\documentclass[aps,prl,preprint,tightenlines,superscriptaddress,showpacs,preprintnumbers,amsmath,amssymb]{revtex4-1}
\def\etagg{\eta_{2\gamma}}
\def\etappp{\eta_{3\pi}}

\newif\ifprd
\prdfalse

\usepackage{graphicx} 
\usepackage{dcolumn}  
\usepackage[left]{lineno}
\usepackage[colorlinks]{hyperref}
\usepackage[hyphenbreaks]{breakurl}

\graphicspath{{ps}}

\renewcommand{\arraystretch}{1.1}

\begin{document}



\title{ \quad\\[0.5cm] Measurement of Time-dependent {\boldmath $CP$} Asymmetries in {\boldmath $B^{0}\to K_S^0 \eta \gamma$} Decays}

\noaffiliation
\affiliation{University of the Basque Country UPV/EHU, 48080 Bilbao}
\affiliation{Beihang University, Beijing 100191}
\affiliation{Brookhaven National Laboratory, Upton, New York 11973}
\affiliation{Budker Institute of Nuclear Physics SB RAS, Novosibirsk 630090}
\affiliation{Faculty of Mathematics and Physics, Charles University, 121 16 Prague}
\affiliation{University of Cincinnati, Cincinnati, Ohio 45221}
\affiliation{Deutsches Elektronen--Synchrotron, 22607 Hamburg}
\affiliation{Duke University, Durham, North Carolina 27708}
\affiliation{University of Florida, Gainesville, Florida 32611}
\affiliation{Key Laboratory of Nuclear Physics and Ion-beam Application (MOE) and Institute of Modern Physics, Fudan University, Shanghai 200443}
\affiliation{Justus-Liebig-Universit\"at Gie\ss{}en, 35392 Gie\ss{}en}
\affiliation{Gifu University, Gifu 501-1193}
\affiliation{II. Physikalisches Institut, Georg-August-Universit\"at G\"ottingen, 37073 G\"ottingen}
\affiliation{SOKENDAI (The Graduate University for Advanced Studies), Hayama 240-0193}
\affiliation{Gyeongsang National University, Chinju 660-701}
\affiliation{Hanyang University, Seoul 133-791}
\affiliation{University of Hawaii, Honolulu, Hawaii 96822}
\affiliation{High Energy Accelerator Research Organization (KEK), Tsukuba 305-0801}
\affiliation{J-PARC Branch, KEK Theory Center, High Energy Accelerator Research Organization (KEK), Tsukuba 305-0801}
\affiliation{IKERBASQUE, Basque Foundation for Science, 48013 Bilbao}
\affiliation{Indian Institute of Technology Bhubaneswar, Satya Nagar 751007}
\affiliation{Indian Institute of Technology Guwahati, Assam 781039}
\affiliation{Indian Institute of Technology Hyderabad, Telangana 502285}
\affiliation{Indian Institute of Technology Madras, Chennai 600036}
\affiliation{Indiana University, Bloomington, Indiana 47408}
\affiliation{Institute of High Energy Physics, Chinese Academy of Sciences, Beijing 100049}
\affiliation{Institute of High Energy Physics, Vienna 1050}
\affiliation{Institute for High Energy Physics, Protvino 142281}
\affiliation{INFN - Sezione di Napoli, 80126 Napoli}
\affiliation{INFN - Sezione di Torino, 10125 Torino}
\affiliation{Advanced Science Research Center, Japan Atomic Energy Agency, Naka 319-1195}
\affiliation{J. Stefan Institute, 1000 Ljubljana}
\affiliation{Kanagawa University, Yokohama 221-8686}
\affiliation{Institut f\"ur Experimentelle Teilchenphysik, Karlsruher Institut f\"ur Technologie, 76131 Karlsruhe}
\affiliation{Kennesaw State University, Kennesaw, Georgia 30144}
\affiliation{King Abdulaziz City for Science and Technology, Riyadh 11442}
\affiliation{Department of Physics, Faculty of Science, King Abdulaziz University, Jeddah 21589}
\affiliation{Korea Institute of Science and Technology Information, Daejeon 305-806}
\affiliation{Korea University, Seoul 136-713}
\affiliation{Kyoto University, Kyoto 606-8502}
\affiliation{Kyungpook National University, Daegu 702-701}
\affiliation{LAL, Univ. Paris-Sud, CNRS/IN2P3, Universit\'{e} Paris-Saclay, Orsay}
\affiliation{\'Ecole Polytechnique F\'ed\'erale de Lausanne (EPFL), Lausanne 1015}
\affiliation{P.N. Lebedev Physical Institute of the Russian Academy of Sciences, Moscow 119991}
\affiliation{Faculty of Mathematics and Physics, University of Ljubljana, 1000 Ljubljana}
\affiliation{Ludwig Maximilians University, 80539 Munich}
\affiliation{Luther College, Decorah, Iowa 52101}
\affiliation{University of Malaya, 50603 Kuala Lumpur}
\affiliation{University of Maribor, 2000 Maribor}
\affiliation{Max-Planck-Institut f\"ur Physik, 80805 M\"unchen}
\affiliation{School of Physics, University of Melbourne, Victoria 3010}
\affiliation{University of Mississippi, University, Mississippi 38677}
\affiliation{University of Miyazaki, Miyazaki 889-2192}
\affiliation{Moscow Physical Engineering Institute, Moscow 115409}
\affiliation{Moscow Institute of Physics and Technology, Moscow Region 141700}
\affiliation{Graduate School of Science, Nagoya University, Nagoya 464-8602}
\affiliation{Kobayashi-Maskawa Institute, Nagoya University, Nagoya 464-8602}
\affiliation{Universit\`{a} di Napoli Federico II, 80055 Napoli}
\affiliation{Nara Women's University, Nara 630-8506}
\affiliation{National United University, Miao Li 36003}
\affiliation{Department of Physics, National Taiwan University, Taipei 10617}
\affiliation{H. Niewodniczanski Institute of Nuclear Physics, Krakow 31-342}
\affiliation{Nippon Dental University, Niigata 951-8580}
\affiliation{Niigata University, Niigata 950-2181}
\affiliation{University of Nova Gorica, 5000 Nova Gorica}
\affiliation{Novosibirsk State University, Novosibirsk 630090}
\affiliation{Osaka City University, Osaka 558-8585}
\affiliation{Pacific Northwest National Laboratory, Richland, Washington 99352}
\affiliation{Panjab University, Chandigarh 160014}
\affiliation{Peking University, Beijing 100871}
\affiliation{University of Pittsburgh, Pittsburgh, Pennsylvania 15260}
\affiliation{Theoretical Research Division, Nishina Center, RIKEN, Saitama 351-0198}
\affiliation{University of Science and Technology of China, Hefei 230026}
\affiliation{Showa Pharmaceutical University, Tokyo 194-8543}
\affiliation{Soongsil University, Seoul 156-743}
\affiliation{Stefan Meyer Institute for Subatomic Physics, Vienna 1090}
\affiliation{Sungkyunkwan University, Suwon 440-746}
\affiliation{Department of Physics, Faculty of Science, University of Tabuk, Tabuk 71451}
\affiliation{Tata Institute of Fundamental Research, Mumbai 400005}
\affiliation{Excellence Cluster Universe, Technische Universit\"at M\"unchen, 85748 Garching}
\affiliation{Department of Physics, Technische Universit\"at M\"unchen, 85748 Garching}
\affiliation{Toho University, Funabashi 274-8510}
\affiliation{Department of Physics, Tohoku University, Sendai 980-8578}
\affiliation{Earthquake Research Institute, University of Tokyo, Tokyo 113-0032}
\affiliation{Department of Physics, University of Tokyo, Tokyo 113-0033}
\affiliation{Tokyo Institute of Technology, Tokyo 152-8550}
\affiliation{Tokyo Metropolitan University, Tokyo 192-0397}
\affiliation{Virginia Polytechnic Institute and State University, Blacksburg, Virginia 24061}
\affiliation{Wayne State University, Detroit, Michigan 48202}
\affiliation{Yamagata University, Yamagata 990-8560}
\affiliation{Yonsei University, Seoul 120-749}
  \author{H.~Nakano}\affiliation{Department of Physics, Tohoku University, Sendai 980-8578} 
  \author{A.~Ishikawa}\affiliation{Department of Physics, Tohoku University, Sendai 980-8578} 
  \author{K.~Sumisawa}\affiliation{High Energy Accelerator Research Organization (KEK), Tsukuba 305-0801}\affiliation{SOKENDAI (The Graduate University for Advanced Studies), Hayama 240-0193} 
  \author{H.~Yamamoto}\affiliation{Department of Physics, Tohoku University, Sendai 980-8578} 
  \author{I.~Adachi}\affiliation{High Energy Accelerator Research Organization (KEK), Tsukuba 305-0801}\affiliation{SOKENDAI (The Graduate University for Advanced Studies), Hayama 240-0193} 
  \author{H.~Aihara}\affiliation{Department of Physics, University of Tokyo, Tokyo 113-0033} 
  \author{S.~Al~Said}\affiliation{Department of Physics, Faculty of Science, University of Tabuk, Tabuk 71451}\affiliation{Department of Physics, Faculty of Science, King Abdulaziz University, Jeddah 21589} 
  \author{D.~M.~Asner}\affiliation{Brookhaven National Laboratory, Upton, New York 11973} 
  \author{V.~Aulchenko}\affiliation{Budker Institute of Nuclear Physics SB RAS, Novosibirsk 630090}\affiliation{Novosibirsk State University, Novosibirsk 630090} 
  \author{T.~Aushev}\affiliation{Moscow Institute of Physics and Technology, Moscow Region 141700} 
% ...
\collaboration{The Belle Collaboration}

\begin{abstract}
We report a measurement of time-dependent $CP$ violation parameters in ${B^0 \to K_S^0 \eta \gamma}$ decays.
The study is based on a data sample, containing ${772 \times 10^6 B\bar{B}}$ pairs, that was collected at the $\Upsilon(4S)$ resonance
with the Belle detector at the KEKB asymmetric-energy $e^+ e^-$ collider.
We obtain the $CP$ violation parameters of
${{\cal S} = -1.32 \pm 0.77 {\rm (stat.)} \pm 0.36{\rm (syst.)}}$ and 
${{\cal A} = -0.48 \pm 0.41 {\rm (stat.)} \pm 0.07{\rm (syst.)}}$ for the invariant mass of the $K_S^0 \eta$ system up to 2.1~GeV/$c^2$.
\end{abstract}

\maketitle

\tighten

{\renewcommand{\thefootnote}{\fnsymbol{footnote}}}
\setcounter{footnote}{0}


\section{INTRODUCTION}


The radiative ${b \to s \gamma}$ decay proceeds dominantly via one-loop electromagnetic penguin diagrams at lowest order in the standard model~(SM).
Since heavy unobserved particles might enter in the loop, such decays are sensitive to new physics~(NP).
Precision measurements of the branching fraction for $B \to X_s \gamma$ by CLEO~\cite{Chen:2001fja}, BaBar~\cite{Aubert:2007my,Lees:2012ym,Lees:2012wg} and Belle~\cite{Limosani:2009qg,Saito:2014das} are consistent with SM predictions~\cite{Becher:2006pu,Misiak:2015xwa} and give a strong constraint to NP models~\cite{BFNP}.
Another important observable that is sensitive to NP signatures in the ${b \to s \gamma}$ process 
is the photon polarization. Within the SM, the photon is mostly produced with left-handed polarization; the right-handed contribution is suppressed by $m_s/m_b$ at leading order, where $m_s$~($m_b$) is the strange~(bottom) quark mass. 
Various NP models, such as supersymmetry~\cite{SUSY1,SUSY2,SUSY3,SUSY4,SUSY5,SUSY6}, left-right symmetric models~\cite{Mohapatra:1974hk} and extra-dimensions~\cite{Randall:1999ee,Goldberger:1999wh,Davoudiasl:1999tf,Pomarol:1999ad,Chang:1999nh,Gherghetta:2000qt}, allow right-handed currents in the loops and hence can enhance the right-handed photon contribution~\cite{Fujikawa:1993zu,Cho:1993zb,Babu:1993hx,Everett:2001yy,Agashe:2004ay}.
Thus, a measurement of the photon polarization in the ${b \to s \gamma}$ process is  an important tool to search for NP.

Several methods have been proposed to measure the photon polarization in the $b \to s \gamma$ process. A measurement of time-dependent $CP$ violation in ${B^0 \to P_1^0 P_2^0 \gamma}$ is the most promising one, where $P_1^0$ and $P_2^0$ are scalar or pseudoscalar mesons and the $P_1^0P_2^0$ system is a $CP$ eigenstate~\cite{Atwood:1997zr,Atwood:2004jj}. 
As the left-~(right-)handed photon contributions are suppressed in $B^0$~($\bar{B}^0$) decays in the SM, an interference between ${\bar{B}^0 \to P_1^0 P_2^0 \gamma_{L(R)}}$ and ${B^0 \to P_1^0 P_2^0 \gamma_{L(R)}}$ can generate a small mixing-induced $CP$ violation parameterized by ${{\cal{S}} \sim -2\xi_{CP}(m_s/m_b) \sin{2\phi_1} \sim -0.02\xi_{CP}}$. Here, $\xi_{CP}$ is the $CP$ eigenvalue of the $P_1^0P_2^0$ system, and $\phi_1$ is an interior angle of the Cabibbo-Kobayashi-Maskawa unitarity triangle~\cite{Cabibbo:1963yz,Kobayashi:1973fv}, defined as ${\phi_1 \equiv {\rm arg}[-V_{cd}V_{cb}^*/V_{td}V_{tb}^*]}$.
Potential contributions from NP-associated right-handed currents could enhance the value of $S$ in the ${B^0 \to P_1^0 P_2^0 \gamma}$ process~\cite{Atwood:1997zr,Chua:1998dx,Chun:2000hj,Chua:2003xq,Goto:2007ee,Ko:2008zu,Blanke:2012tv,Shimizu:2012zw,Kou:2013gna,Haba:2015gwa,Malm:2015oda}.

At Belle and BaBar, the $CP$ violation parameters for the ${b \to s \gamma}$ transition were measured in the decays of ${B^0 \to K_S^0 \pi^0 \gamma}$ including ${K^{*0} \to K_S^0 \pi^0}$~\cite{Ushiroda:2006fi,Aubert:2008gy}, ${B^0 \to K_S^0 \eta \gamma}$~\cite{Aubert:2008js}, ${B^0 \to K_S^0 \rho^0 \gamma}$~\cite{Li:2008qma,Sanchez:2015pxu}, and ${B^0 \to K_S^0 \phi \gamma}$~\cite{Sahoo:2011zd}. All results are consistent with the SM prediction within the uncertainties~\cite{Grinstein:2004uu,Grinstein:2005nu,Matsumori:2005ax,Ball:2006cva,Ball:2006eu,Jager:2014rwa}. In this paper, we report the first measurement of time-dependent $CP$ violation in ${B^0 \to K_S^0 \eta \gamma}$ at Belle. The study is based on the full data sample of 711~fb$^{-1}$ containing ${772\times 10^6 B\bar{B}}$ pairs recorded at the $\Upsilon(4S)$ resonance with the Belle detector~\cite{Belle} at the KEKB $e^+ e^-$ collider~\cite{KEKB}.


\section{Time-Dependent $CP$ Violation}

At the KEKB asymmetric-energy collider (3.5 GeV $e^+$ on 8.0 GeV $e^-$), 
the $\Upsilon(4S)$ is produced with a Lorentz boost of $\beta\gamma=0.425$ 
nearly along the $z$ axis, which is antiparallel to the $e^+$ beam direction.
In the decay chain ${\Upsilon(4S) \to B^0\bar{B}^0 \to f_{\rm rec}f_{\rm tag}}$,
one of the $B$ mesons decays at proper time $t_{\rm rec}$ to a final
state $f_{\rm rec}$~(our signal mode), and the other ($B_{\rm tag}$) decays
at proper time $t_{\rm tag}$ to a final state $f_{\rm tag}$ that is used to determine
the flavor of the signal $B$ meson. The distribution of the proper time difference
$\Delta t = t_{\rm rec} - t_{\rm tag}$ is given by
\ifprd
\begin{eqnarray}
 {\cal{P}}(\Delta t) = \frac{e^{-|\Delta t|/\tau_{B^0}}}{4\tau_{B^0}} \left\{ 1 \right. &+& q \left[{\cal{S}} \sin(\Delta m_d \Delta t) \right. \\
&+& \left. \left. {\cal{A}}\cos(\Delta m_d \Delta t) \right] \right\},
    \label{eq:deltat}
\end{eqnarray}
\else
\begin{eqnarray}
 {\cal{P}}(\Delta t) = \frac{e^{-|\Delta t|/\tau_{B^0}}}{4\tau_{B^0}} \left\{ 1 \right. &+& q \left[{\cal{S}} \sin(\Delta m_d \Delta t) \right. + \left. \left. {\cal{A}}\cos(\Delta m_d \Delta t) \right] \right\},
    \label{eq:deltat}
\end{eqnarray}
\fi
where ${\cal{S}}$~(${\cal{A}}$) is the mixing-induced~(direct) $CP$ violation parameter,
$\tau_{B^0}$ is the $B^0$ lifetime, $\Delta m_d$ is the mass difference
between the two $B^0$ mass eigenstates, and $q = +1$ ($-1$) is the $b$-flavor charge when the
tagging $B$ meson is a $B^0$ ($\bar{B}^0$).
Since the $B^0$ and $\bar{B}^0$ mesons are approximately at 
rest in the $\Upsilon(4S)$ center-of-mass (CM) frame,
$\Delta t$ can be determined from the displacement in $z$ in the laboratory frame
between the $f_{\rm rec}$ and $f_{\rm tag}$ decay vertices:
${\Delta t \simeq (z_{\rm rec} - z_{\rm tag})/\beta\gamma c  \equiv \Delta z/\beta\gamma c}$,
where $z_{\rm rec}$ and $z_{\rm tag}$ are the decay positions along the $z$ axis of the signal 
and tag-side $B$ mesons.

\section{Belle Detector}

The Belle detector~\cite{Belle} is a large-solid-angle magnetic
spectrometer that
consists of a silicon vertex detector (SVD),
a 50-layer central drift chamber (CDC), an array of
aerogel threshold Cherenkov counters (ACC), 
a barrel-like arrangement of time-of-flight
scintillation counters (TOF), and an electromagnetic caloriemeter
(ECL) comprised of CsI(Tl) crystals.
All these detector components are located inside 
a superconducting solenoid coil that provides a 1.5~T
magnetic field.  An iron flux-return located outside of
the coil is instrumented with resistive plate chambers to detect $K_L^0$ mesons and muons. 
Two inner detector configurations were used: A 2.0 cm radius beampipe
and a 3-layer SVD was used for the first sample
of ${152\times 10^6 B\bar{B}}$ pairs~(SVD1), while a 1.5 cm radius beampipe, a 4-layer
SVD and a small-inner-cell CDC were used to record  
the remaining ${620\times 10^6 B\bar{B}}$ pairs~(SVD2)~\cite{svd2}.  

\section{Event Selection}

The most energetic isolated cluster in the ECL in the CM frame of an event that is not associated 
with any charged tracks reconstructed in the SVD and CDC is selected as the prompt photon.
Its energy must lie between 1.8 and 3.4~GeV.
We require that its shower shape be consistent with an electromagnetic shower
by imposing the criterion $E_{9}/E_{25}$ $>$ 0.95 for the ratio of energy 
deposits in a 3 $\times$ 3 array of CsI(Tl) crystals 
to that in a 5 $\times$ 5 array, both centered on the crystal with the largest energy deposit.
To reduce contamination from the decays 
${\pi^{0} \to \gamma \gamma}$ or ${\eta \to \gamma \gamma}$, 
the prompt photon candidate is paired with all other photons
in the event with energy exceeding 40~MeV in the laboratory frame. We reject
the event if the pair is consistent with the above decays,
based on a likelihood constructed from the invariant mass, 
the energy and polar angle of the second photon in the laboratory frame~\cite{Koppenburg:2004fz}.

Neutral pion candidates are reconstructed from two photons whose energies 
exceed 50~MeV in the laboratory frame. We require the invariant mass of the photon pairs to lie between 114 and 147~MeV$/c^2$, which corresponds approximately to a $\pm3\sigma$ window in resolution about the nominal $\pi^0$ mass~\cite{Patrignani:2016xqp}.
To reduce the combinatorial background, we retain candidates with a momentum greater than 100~MeV$/c$ in the CM frame.

Charged particles, except for pions from $K_S^0$ decays, are required to have a distance 
of closest approach to the interaction point~(IP) within 5.0~cm along the $z$ axis and 0.5~cm 
in the transverse plane. Charged kaons and pions are identified with a likelihood ratio 
constructed from specific ionization measurements in the CDC, time-of-flight information from the TOF, 
and the number of photoelectrons in the ACC. 

Neutral kaon ($K_S^0$) candidates are reconstructed from pairs of oppositely-charged tracks, treated as pions, and identified by a
multivariate analysis~\cite{NB} based on two sets of input variables~\cite{nakano}. 
The first set that separates $K_S^0$ candidates from the combinatorial background are:
(1) the $K_S^0$ momentum in the laboratory frame,
(2) the distance along the $z$ axis between the two track helices at their closest approach,
(3) the flight length in the $x$-$y$ plane,
(4) the angle between the $K_S^0$ momentum and the vector joining its decay vertex to the nominal IP,
(5) the angle between the $\pi$ momentum and the laboratory-frame direction of the $K_S^0$ in its rest frame,
(6) the distances of closest approach in the $x$-$y$ plane between the IP and the pion helices,
(7) the numbers of hits for axial and stereo wires in the CDC for each pion,
and (8) the presence or absence of associated hits in the SVD for each pion.
The second set of variables, which identifies ${\Lambda \to p\pi^-}$ background that has a similar long-lived topology, are:
(1) particle identification information, momentum, and polar angles of the two daughter tracks in the laboratory frame,
and (2) the invariant mass calculated with the proton- and pion-mass hypotheses for the two tracks.
In total, the first and second sets comprise 13 and 7 input variables, respectively. 
The selected $K^0_S$ candidates are required to have an invariant mass within $\pm10$~MeV/$c^2$ of
the nominal value, corresponding to a $\pm$3$\sigma$ interval in mass resolution.


We reconstruct $\eta$ candidates from the $\gamma \gamma$ and $\pi^{+}\pi^{-}\pi^{0}$ final states, denoted as $\etagg$ and $\etappp$, respectively. For the $\etagg$ mode, 
we require that the photon energy in the CM system be greater than 150 MeV.
The candidates satisfying the di-photon invariant mass requirement of ${510~{\rm MeV}/c^2 < M_{\gamma\gamma} < 575~{\rm MeV}/c^2}$
are retained.
For the $\etappp$ mode, 
the invariant mass of the three-pion system is required to be in the range 537~MeV$/c^2$ $<$ $M_{\pi\pi\pi}$ $<$ 556~MeV$/c^2$.
These requirements correspond to about $\pm2\sigma$ windows in mass resolution. 

We reconstruct $B$ candidates by combining a $K_S^0$ with an $\eta$ and a $\gamma$ candidate. 
We form two kinematic variables to select $B$ mesons:
the energy difference ${\Delta E \equiv E_{B}^{\rm CM} - E_{\rm beam}^{\rm CM}}$
and the beam-energy constrained mass ${M_{\rm bc} \equiv  \sqrt{ (E_{\rm beam}^{\rm CM}/c^2)^2 - (p_{B}^{\rm CM}/c)^2 }}$,
where $E_{\rm beam}^{\rm CM}$ is the beam energy, and $E_{B}^{\rm CM}$ and $p_{B}^{\rm CM}$ are 
the energy and momentum, respectively, of the $B$ candidate in the CM system.
We define the signal region in $\Delta E$ and $M_{\rm bc}$ for the measurement of $CP$ violation as $-0.15$~GeV $ < \Delta E <$ 0.08~GeV and 5.27~GeV$/c^2$ $< M_{\rm bc} <$ 5.29~GeV$/c^2$. To determine the signal fraction, a larger fitting region, ${|\Delta E| < 0.5}$~GeV and 5.20~GeV$/c^2$ $< M_{\rm bc} <$ 5.29~GeV$/c^2$, is employed.
The average number of $B$ candidates in an event with at least one candidate is 1.47; this is primarily due to multiple $\eta$ candidates.
If there is more than one $B$ candidate in the fitting region, the candidate whose $\eta$ daughter's mass is closest to the nominal value is selected. If still necessary, the $B$ candidate with the $K_S^0$ daughter's mass closest to the nominal value is retained.


\section{Background Suppression}
To suppress the dominant ${e^+ e^- \to q\bar{q}}$ ($q$ $\in$ \{$u$,\,$d$,\,$s$,\,$c$\}) continuum background, we use a neural network
based on four input variables calculated in the CM frame: 
(1) the cosine of the angle between the $B$ momentum and the $z$ axis, 
(2) the likelihood ratio of modified Fox-Wolfram moments~\cite{SFW,KSFW} that gives the strongest separation power, 
(3) the cosine of the angle between the third sphericity axes~\cite{sph} calculated from the $B$ candidate and all other particles in the rest of the event~(ROE), and 
(4) the cosine of the angle between the first sphericity axis in the ROE and the $z$ axis. 
The network is trained with a GEANT3-based Monte Carlo~(MC) simulation~\cite{GEANT}.
The output variable, ${\cal{O}}_{\rm NB}$, in the range [$-1$, $1$], is used as one of the variables to determine the signal fraction. 
To enable a simple analytical modeling, ${\cal{O}}_{\rm NB}$ is transformed into
\begin{eqnarray}
 {\cal{O}}_{\rm NB}' = \ln{\frac{ {\cal{O}}_{\rm NB} - {\cal{O}}_{\rm NB}^{\rm min} }{ {\cal{O}}_{\rm NB}^{\rm max} - {\cal{O}}_{\rm NB}} },
    \label{eq:nb}
\end{eqnarray}
where ${\cal{O}}_{\rm NB}^{\rm min}$ and ${\cal{O}}_{\rm NB}^{\rm max}$ are chosen to be $-0.7$ and $0.935$~($0.915$), respectively, for the $\etagg$ ($\etappp$) mode.
The events with ${\cal{O}}_{\rm NB} < {\cal{O}}_{\rm NB}^{\rm min}$ are discarded; this selection keeps 80\%~(73\%) of the signal while removing 92\%~(95\%) $q\bar{q}$ background for the $\etagg$~($\etappp$) mode.

The decay modes of the following $CP$ eigenstates constitute peaking backgrounds:
${B^{0} \to J/\psi (\eta \gamma) K_S^0}$, ${B^0 \to a_X (\eta \pi^0) K_S^0}$, ${B^{0}\to \bar{D}^0 (K_S^0 \eta) \pi^0}$, ${B^{0}\to \bar{D}^0 (K_S^0 \eta) \eta}$, ${B^{0}\to \bar{D}^0 (K_S^0 \pi^0) \eta}$, and ${B^0 \to \eta K_X (K_S^0 \pi^0)}$, where $a_X$ and $K_X$ represent a light unflavored resonance and a kaonic resonance, respectively.
To suppress these backgrounds, we require 
2.0~GeV$/c^2$ $< M_{\gamma \eta} <$ 2.9~GeV$/c^2$ or $M_{\gamma \eta} >$ 3.2~GeV$/c^2$ to eliminate ${J/\psi \to \eta \gamma}$ and ${a_X \to \eta \pi^0}$,
$M_{K \eta} <$~1.82~GeV$/c^2$ or $M_{K \eta} >$ 1.90~GeV$/c^2$ to remove ${\bar{D}^0 \to K_S^0 \eta}$, 
and $M_{\gamma K} >$ 2.0~GeV$/c^2$ to suppress ${K_X \to K_S^0 \pi^0}$ and ${\bar{D}^0 \to K_S^0 \pi^0}$,
where a soft photon from the $\pi^0$ decay is undetected.

One of the decays arising from the $b \to s \gamma$ transition, ${B^0 \to K_S^0 \pi^0 \gamma}$, is a major peaking background.
This decay is exclusively reconstructed and rejected if the candidate is found to satisfy the following requirements:
$0.12$~GeV$/c^2$ $<$ $M_{\gamma\gamma}$ $<$ $0.15$~GeV$/c^2$, 
$1.6$~GeV $<$ $E_{\gamma}^{\rm CM}$ $<$ $3.4$~GeV, 
$-0.20$~GeV $<$ $\Delta E$ $<$ $0.10$~GeV, and 
$M_{\rm bc} >$~5.27~GeV$/c^2$.

\section{Helicity Angle and Mass Distributions}

\begin{figure}[h]
        \begin{center}
\ifprd        
        \includegraphics[width=0.2375\textwidth]{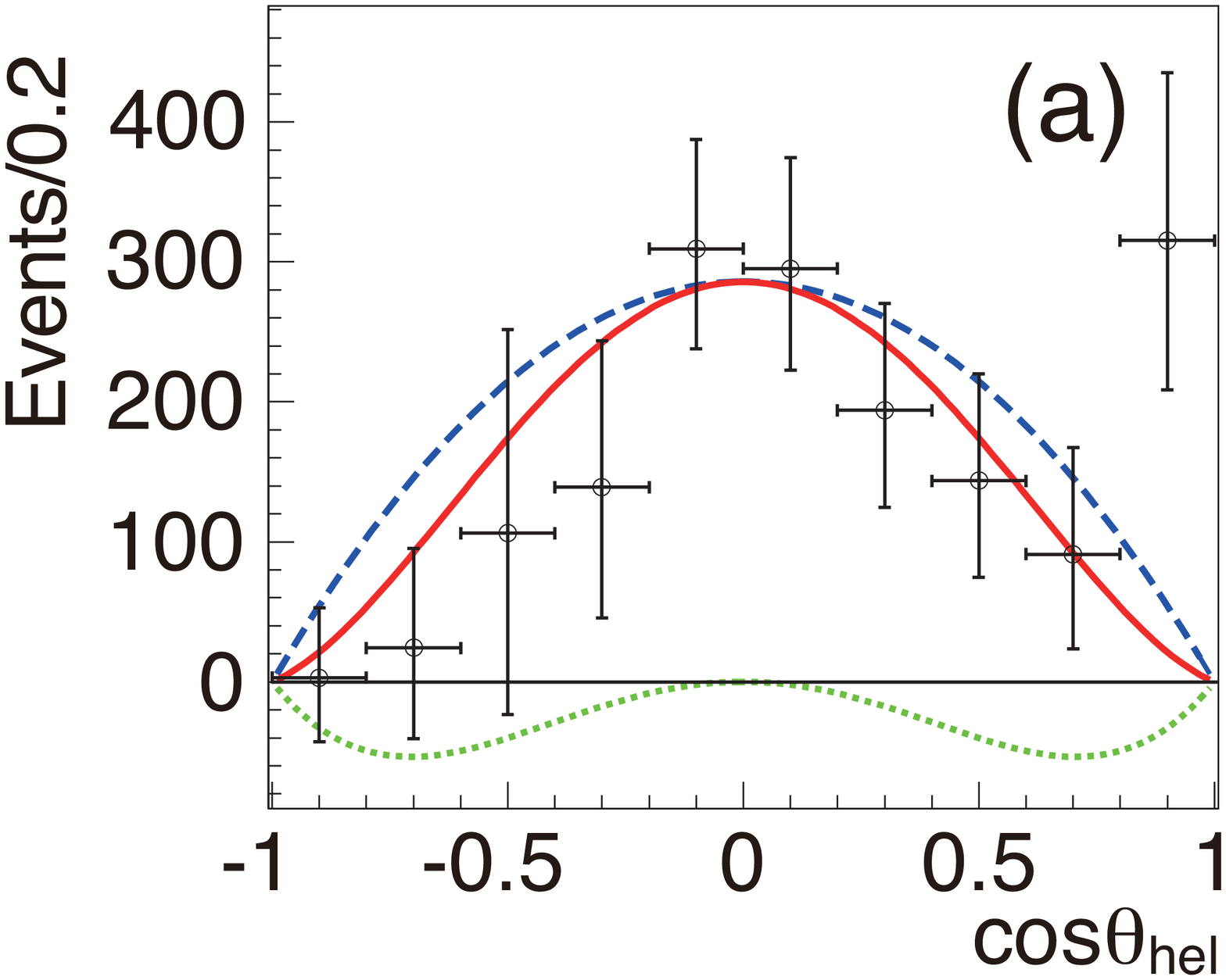}
        \includegraphics[width=0.2375\textwidth]{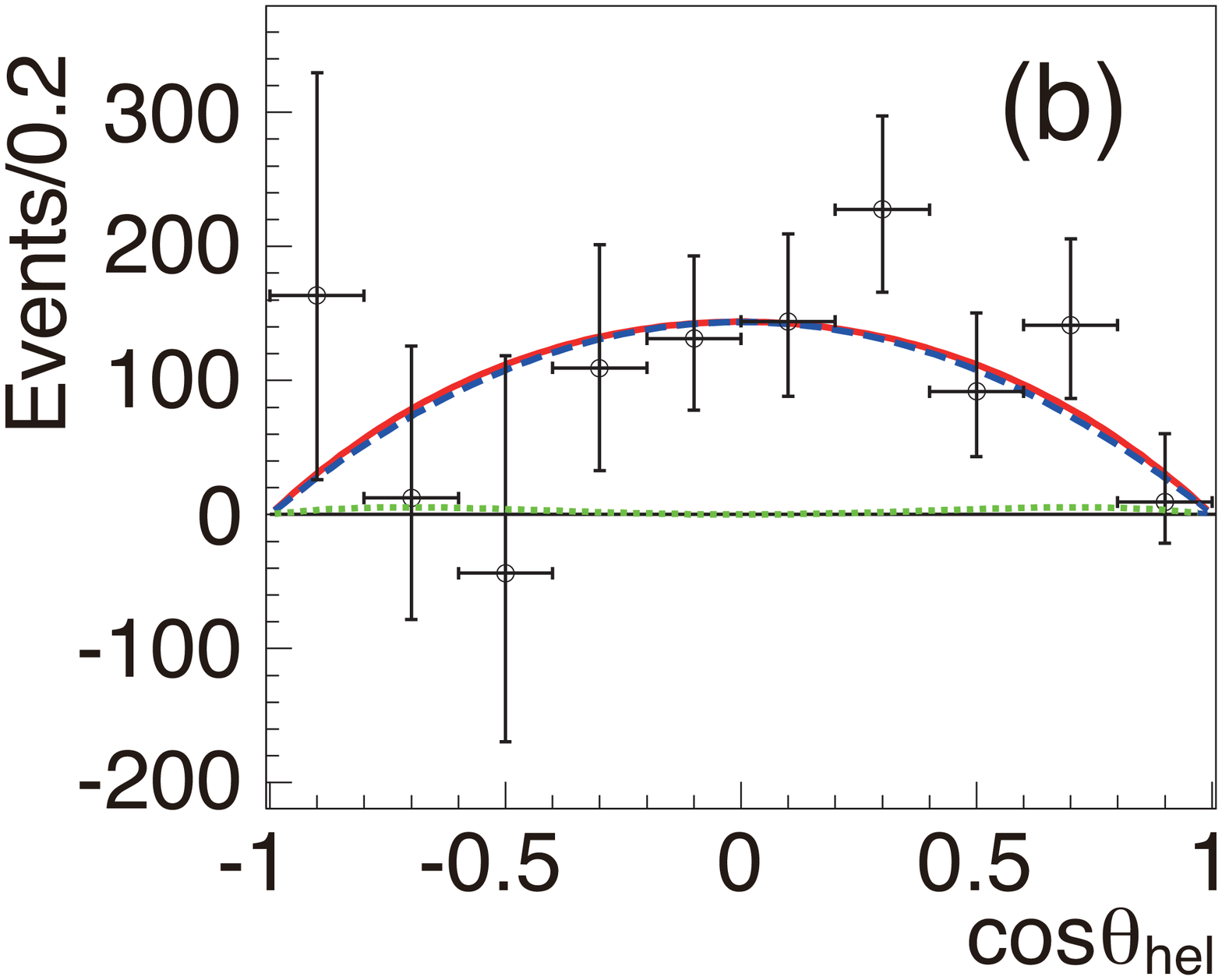}
\else
        \includegraphics[width=0.4\textwidth]{fig1A.eps}
        \includegraphics[width=0.4\textwidth]{fig1B.eps}
\fi
        \end{center}
    \caption {Background-subtracted and efficiency-corrected helicity angle distributions of $B^+ \to K^+\eta\gamma$ for (a) $\etagg$ and (b) $\etappp$ modes. The solid red curve shows the fit result, the dashed blue curve is the spin-1 component, and the dotted green line is the spin-2 component.
             }
    \label{fig:distribution_h}
\end{figure}

\begin{figure}[h]
\begin{center}
\ifprd
        \includegraphics[width=0.2375\textwidth]{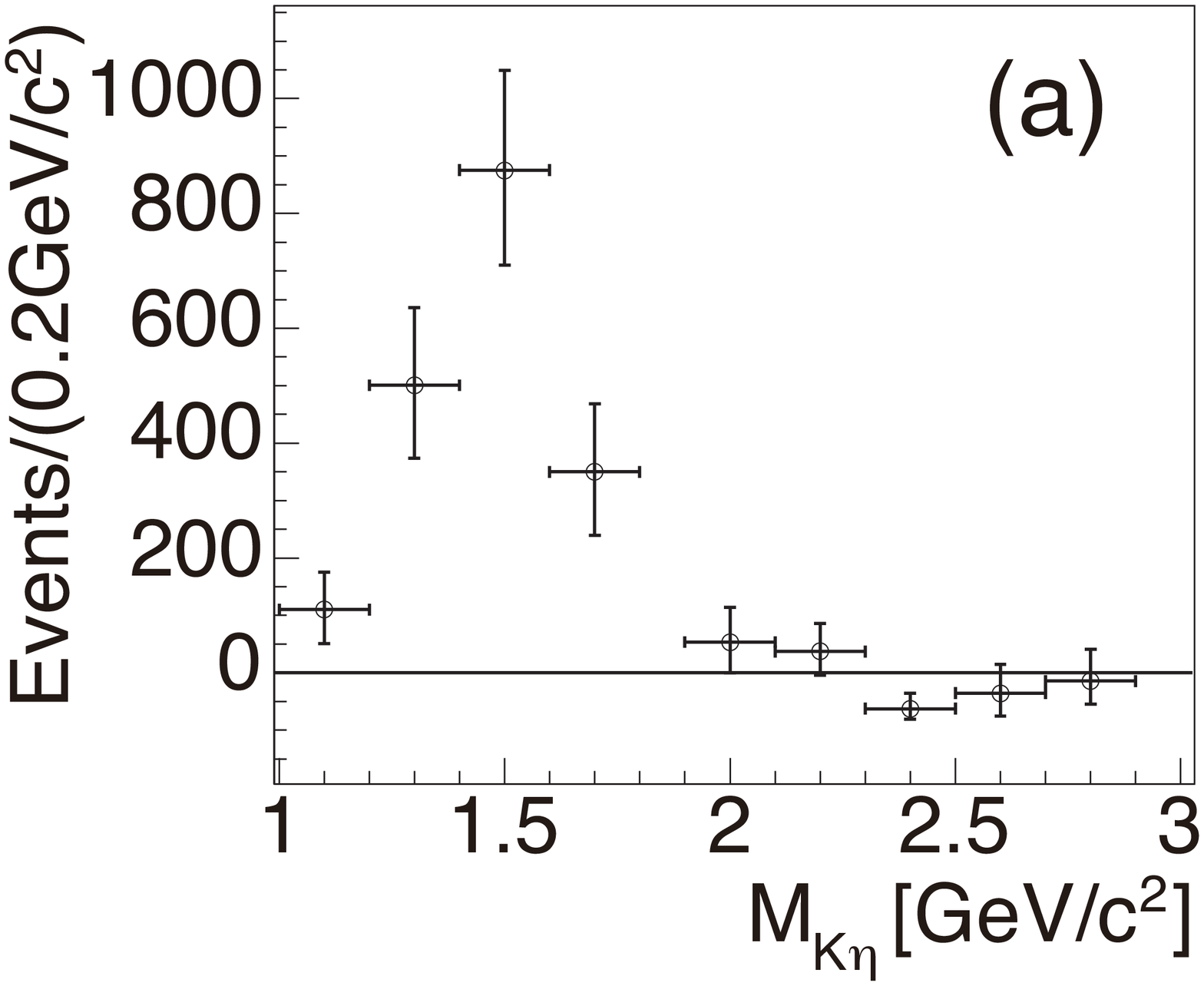}
        \includegraphics[width=0.2375\textwidth]{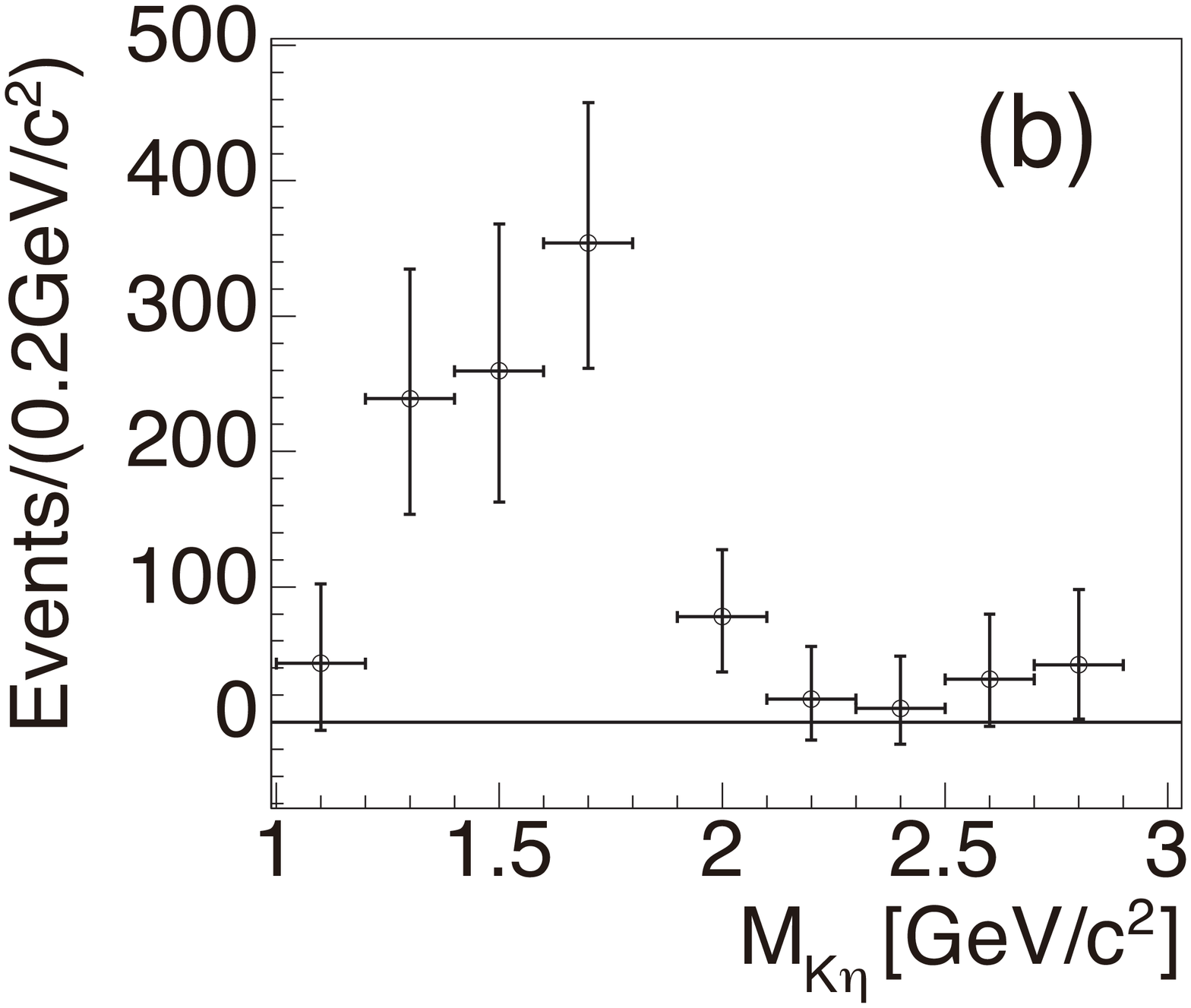}
\else
        \includegraphics[width=0.4\textwidth]{fig2A.eps}
        \includegraphics[width=0.4\textwidth]{fig2B.eps}
\fi
\end{center}
    \caption {Background-subtracted and efficiency-corrected invariant mass distributions of the $K^+ \eta$ system for the (a) $\etagg$ and (b) $\etappp$ modes.}
    \label{fig:distribution_m}
\end{figure}

As the spin and invariant mass of the $K\eta$ system are not well known,
we study $B^+ \to K^+ \eta \gamma$~\cite{CC} assuming the isospin symmetry breaking to be small between $B^0 \to K^0 \eta \gamma$ and $B^+ \to K^+ \eta \gamma$~\cite{isospin}.
The selections on $B^+ \to K^+ \eta \gamma$ are the same as those on $B^0 \to K_S^0 \eta \gamma$ except for kaon selections.
We define the helicity angle~($\theta_{\rm hel}$) as 
the angle between the $K$ momentum and the
opposite of the $B$-meson momentum in the $K\eta$ rest frame.
The signal yields are extracted by fitting to $\Delta E$ and $M_{\rm bc}$ in bins of
$\cos\theta_{\rm hel}$ and the $K^+\eta$ invariant mass; later, the efficiency-corrected
yield is obtained.
We fit to the $\cos\theta_{\rm hel}$ distribution with spin-1 and spin-2 hypotheses, as
a spin-3 resonance in $B$ decays is only found in a $B_s^0$ decay and is highly suppressed compared to the
spin-1 states~\cite{Aaij:2014xza}.
Figures~\ref{fig:distribution_h} and \ref{fig:distribution_m} show the background-subtracted and efficiency-corrected $\theta_{\rm hel}$
and invariant-mass distributions for $B^+ \to K^+\eta\gamma$.
We find that the signal is concentrated in the region $M_{K\eta} < 2.1$~GeV$/c^2$ and has the signature of a spin-1 system.
From these studies, we apply two selection criteria, $-0.7 < \cos\theta_{\rm hel} < 0.9$ and
$M_{K\eta}$ $<$ 2.1~GeV$/c^2$, to $B^0 \to K^0_S \eta \gamma$ candidates to maximize the signal sensitivity.

\section{Signal Extraction}

\begin{figure}[htbp]
    \begin{center}
\ifprd
    \includegraphics[width=0.2375\textwidth]{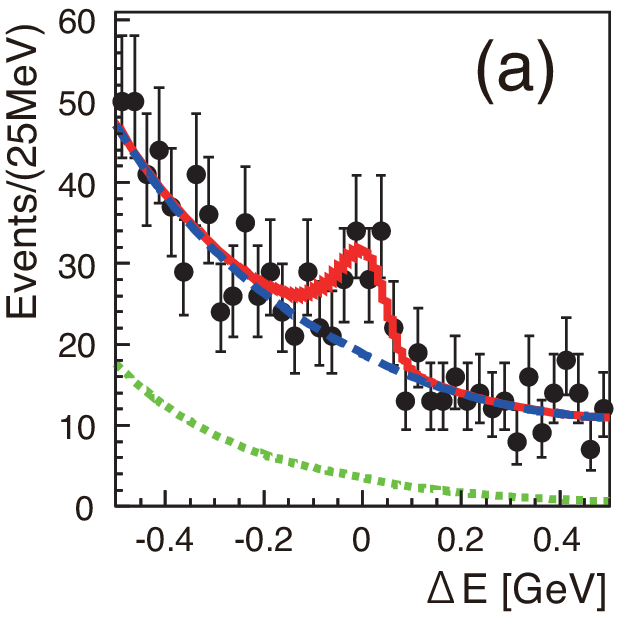}
    \includegraphics[width=0.2375\textwidth]{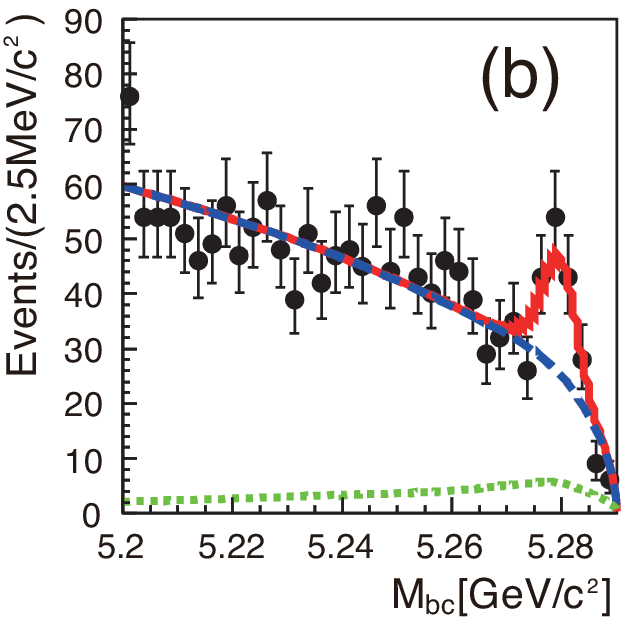}
    \includegraphics[width=0.2375\textwidth]{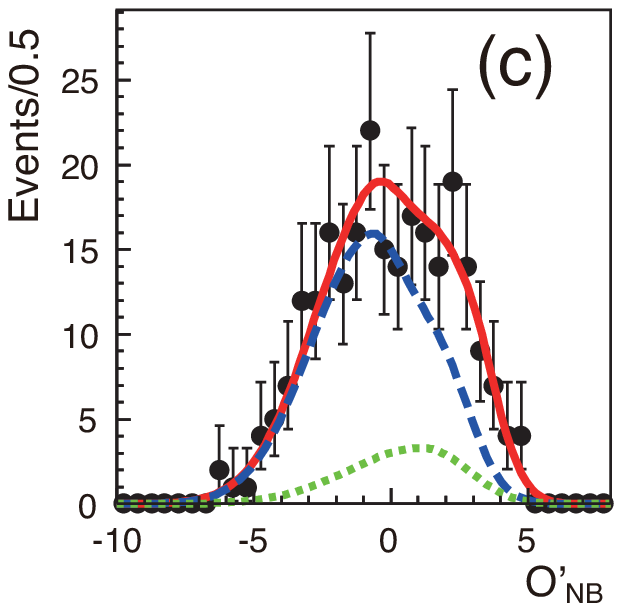}
\else
    \includegraphics[width=0.4\textwidth]{fig3A.eps}
    \includegraphics[width=0.4\textwidth]{fig3B.eps}
    \includegraphics[width=0.4\textwidth]{fig3C.eps}
\fi
    \caption { Projections of the three-dimensional fit onto: (a)~$\Delta E$ in the $M_{\rm bc}$ signal region, (b)~$M_{\rm bc}$ in the $\Delta E$ signal region and (c)~${\cal{O}}_{\rm NB}'$ in the $\Delta E$ and $M_{\rm bc}$ signal regions.
               The solid red curves show the fit results,
               the dotted green curves represent $B\bar{B}$ background, and
               the dashed blue curves describe the total background.
             }
    \label{fig:N_all_tot}
    \end{center}
\end{figure}

We extract the signal yield with a three-dimensional extended unbinned maximum-likelihood fit to $\Delta E$, $M_{\rm bc}$, and ${\cal{O}}_{\rm NB}'$.
For the signal $\Delta E$--$M_{\rm bc}$ distribution, a two-dimensional histogram is used as the two variables
have 40\% correlation due to the imperfect energy measurement for the prompt photon.
The ${\cal{O}}_{\rm NB}'$ distribution is modeled with the sum of two bifurcated Gaussian functions sharing a common peak position and right-side width.
For the $q\bar{q}$ background, the $\Delta E$ and $M_{\rm bc}$ distributions are parameterized by a second-order Chebyshev polynomial
and an ARGUS function~\cite{ARGUS}, respectively.
The sum of a bifurcated Gaussian and a Gaussian function reproduces its ${\cal{O}}_{\rm NB}'$ distribution.
For background from $B$ meson decays, the $\Delta E$ distribution is described by an exponential function;
${\cal{O}}_{\rm NB}'$ is modeled with a bifurcated Gaussian function;
the $M_{\rm bc}$ distribution is described by the sum of an ARGUS function and a Gaussian function.
The fit results projected onto $\Delta E$, $M_{\rm bc}$ and ${\cal{O}}_{\rm NB}'$
are shown in Fig.~\ref{fig:N_all_tot}.
We obtain $69.5^{+13.4}_{-12.4}$ and $22.4^{+7.3}_{-6.4}$ signal events for the $\etagg$ and $\etappp$ decay modes, respectively, 
with purities in the signal region of $28.4\%$ and $22.5\%$.

\section{Flavor Tagging}

The flavor of the $B_{\rm tag}$ meson is determined from inclusive properties of particles in the ROE based on a multi-dimensional likelihood method.
The algorithm for flavor tagging is described in detail elsewhere~\cite{TaggingNIM}.
Two parameters, $q$ defined in Eq.~(\ref{eq:deltat}) and $r$, are used to represent the tagging information.
The parameter $r$ is an event-by-event MC-determined flavor tagging quality factor that ranges from 0 for no flavor
information to 1 for unambiguously determined flavor. The data are sorted into seven intervals of $r$ in which the fractions of wrongly tagged $B$ flavor ($w_l$, $l=1,...,7$) as well as the differences between $B^0$ and $\bar{B}^0$~($\Delta w_l$) are determined from self-tagged semileptonic and hadronic $b \to c$ decays. The total effective tagging efficiency, $\Sigma [f_l \times (1-2w_l)^2]$, where $f_l$ is the fraction of events in category $l$, is determined to be $(29.8 \pm 0.4)$\%.

\section{Vertex Reconstruction}

The vertex positions of signal-side decays of ${B^0 \to K_S^0 \etappp \gamma}$ and ${B^0 \to K_S^0 \etagg \gamma}$ 
is determined from the charged tracks. 
For ${B^0 \to K_S^0 \etappp \gamma}$ decays, we require at least one of the charged pions from $\etappp$ decays, which originate from the $B$ decay position, to have at least one (two) hit in the SVD $r$-$\phi$~($z$) layers. 
To improve the $B$-vertex resolution, we use an additional constraint from the transverse-plane beam profile at the IP ($\sigma_x^{\rm beam} \sim 100$~$\mu$m, $\sigma_y^{\rm beam} \sim 5$~$\mu$m)
smeared with the finite flight length of the $B^0$ meson in the $x$-$y$ plane. The estimated uncertainty of the reconstructed vertex position in the $z$ direction~($\sigma_z^{\rm rec}$) determined with single (two) charged track is required to be less than 500~$\mu$m~(200~$\mu$m) to ensure enough quality for time dependent analysis.
For ${B^0 \to K_S^0 \etagg \gamma}$ decays, the $K_S^0$ trajectory, reconstructed from its pion daughters, is used to determine the vertex position with the aforementioned constraint on the smeared beam profile; this strategy is adopted since the decay vertex of the long-lived $K_S^0$ is displaced from the $B$ decay vertex. To have good resolution of the $K_S^0$ trajectory, both pions daughters must satisfy SVD-hit requirements of at least one (two) hit in the $r$-$\phi$~($z$) layers for SVD1, and at least two hits in both $r$-$\phi$ and $z$ layers for SVD2. We apply a selection on the $\sigma_z^{\rm rec}$ to be less than 500~$\mu$m.
The vertex position of $B_{\rm tag}$ is determined from well-reconstructed charged particles in the ROE~\cite{Chen:2005dra}. 
The $|\Delta t|$ is restricted to be less than 70~ps for further analysis. 

\section{Event Model}

We determine ${\cal S}$ and ${\cal A}$ by performing an unbinned maximum-likelihood fit 
to the observed $\Delta t$ distribution in the signal region.
The probability density function~(PDF) expected for the signal distribution, ${\cal{P}}_{\rm sig}(\Delta t, q, w_l, \Delta w_l; {\cal{S}}, {\cal{A}})$, is given by Eq.~(\ref{eq:deltat}), modified to incorporate the effect of incorrect flavor assignment. Two of the parameters in the PDF expression, $\tau_{B^0}$ and $\Delta m_d$, are fixed to their world average~\cite{Beringer:1900zz}. 
The distribution is convolved with the proper-time resolution function, $R_{\rm sig}(\Delta t)$, which is a function of the event-by-event $\Delta t$ uncertainties.
The resolution function $R_{\rm sig}(\Delta t)$ incorporates the
detector resolution, 
contamination of non-primary tracks in the vertex reconstruction of $B_{\rm tag}$, and 
the kinematic energy generated by the $\Upsilon(4S)$ decay.
As in Ref.~\cite{Sumisawa:2005fz}, universal $R_{\rm sig}$ parameters are used for the vertex reconstruction for $\etappp$ and the long-lived $K_S^0$.
A detailed description can be found in Ref.~\cite{vertexres}.
The PDF for $B\bar{B}$ background events
(${\cal{P}}_{B\bar{B}}$) is modeled in the same way as for signal, but with
different lifetime and $CP$ violation parameters while using the same
resolution function ($R_{B\bar{B}}=R_{\rm sig}$). The effective
lifetime of the $B\bar{B}$ background is obtained from a fit to
the 
MC sample for each $\eta$ decay mode.
The PDF for $q\bar{q}$ background events, ${\cal{P}}_{q\bar{q}}$, is modeled
as the sum of exponential and prompt components, and
is convolved with a double Gaussian representing
the resolution function $R_{q\bar{q}}$. All parameters in ${\cal{P}}_{q\bar{q}}$ and
$R_{q\bar{q}}$ are determined by a fit to the $\Delta t$ distribution of a
background-enhanced sample in the $\Delta E$--$M_{\rm bc}$ sideband.

For each event $i$, the following likelihood function is calculated:
\begin{eqnarray}
    \nonumber
    P_i &=& (1-f_{\rm ol}) \int \Big[ f_{\rm sig}{\cal{P}}_{\rm sig}(\Delta t')R_{\rm sig}(\Delta t_i - \Delta t') \\
    \nonumber
        &+& f_{B\bar{B}}{\cal{P}}_{B\bar{B}}(\Delta t')R_{B\bar{B}}(\Delta t_i - \Delta t')  \\
    \nonumber
        &+& (1-f_{\rm sig}-f_{B\bar{B}}){\cal{P}}_{q\bar{q}}(\Delta t')R_{q\bar{q}}(\Delta t_i - \Delta t') \Big]d\Delta t' \\
        &+& f_{\rm ol}{\cal{P}}_{\rm ol}(\Delta t_i),
\end{eqnarray}
where ${\cal{P}}_{\rm ol}$ is a broad Gaussian function that represents an outlier component with a small fraction $f_{\rm ol}$~\cite{vertexres}.
The signal and background probabilities, $f_{\rm sig}$ and $f_{B\bar{B}}$, are calculated on an event-by-event basis from the function obtained by the same $\Delta E$--$M_{\rm bc}$--${\cal{O}}_{\rm NB}'$ fit used to extract the signal yield, and are then multiplied by a
factor that depends on the flavor tagging $r$-bin. The $r$
distributions of the signal and the $q\bar{q}$ background are estimated
by repeating the $\Delta E$--$M_{\rm bc}$--${\cal{O}}_{\rm NB}'$ fit procedure for each
$r$ interval with the three background shape parameters
fixed to the full-range result. The $B\bar{B}$ background distribution
is estimated from MC samples and found to be small.

\section{Results}

\begin{figure}[htbp]
    \vspace{1cm}
    \begin{center}
\ifprd
    \includegraphics[width=0.48\textwidth]{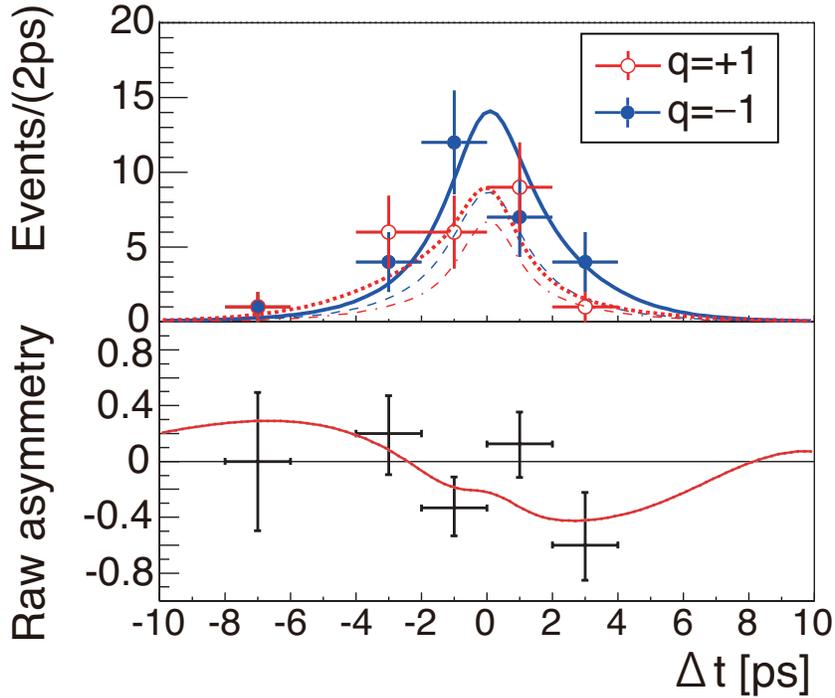}
\else
    \includegraphics[width=0.67\textwidth]{fig4.eps}
\fi
    \end{center}
    \caption { $\Delta t$ distribution (top) and raw asymmetry (bottom) for events in the $0.5 < r \leq 1.0$ region.
               (Top) The filled blue dots show the distribution of $\bar{B}^0$ tagged events and the open red dots show the distribution for $B^0$ tagged events. 
               The solid blue and dotted red curves show the total PDF for $\bar{B}^0$ and $B^0$ tagged events, respectively. The dashed blue and dot-dashed red curves represent the background PDF for $\bar{B}^0$ and $B^0$ tagged events, respectively.
               (Bottom) The solid red curve shows the result of the extended unbinned maximum-likelihood fit.
             }
    \label{fig:CP_goodqr_N_all}
\end{figure}

The only free parameters in the final fit are ${\cal{S}}$ and ${\cal{A}}$,
which are determined by maximizing the likelihood function $L = \Pi_i P_i (\Delta t_i; {\cal S}, {\cal A})$, where the product is over all events. We obtain

\begin{eqnarray}
    \nonumber
    {\cal S} = -1.32 \ \ {\rm and} \ \ {\cal A} = -0.48,
\end{eqnarray}
and find that the central values are outside of the physical boundary defined by ${{\cal S}^2+{\cal A}^2 = 1}$. 
We extract the statistical uncertainties from the root-mean-square of the $CP$ violation parameter distributions obtained using an ensemble test with input values of $( {\cal S}_{{\rm true}}, {\cal A}_{{\rm true}} )=(-0.94, -0.34)$, which is the closest point on the physical boundary to the fit result~\cite{Abe:2003jaa}, as $\delta {\cal S} = \pm 0.77$ and $\delta {\cal A} = \pm 0.41$~\cite{toyMC0}.
The correlation between ${\cal S}$ and ${\cal A}$ is found to be 0.15.
We define the raw asymmetry in each $\Delta t$ interval as
$(N_{q=+1} - N_{q=-1})/(N_{q=+1} + N_{q=-1})$, where $N_{q=\pm1}$ is the number of observed candidates with the given $q$. The $\Delta t$ distributions and raw asymmetries for events in the signal-enhanced $0.5 < r \leq 1.0$ region for $q=\pm1$ are shown in Fig.~\ref{fig:CP_goodqr_N_all}.

\section{Validations}

Various cross-checks are performed to confirm the validity of our procedure.
The $CP$ asymmetry fit to MC signal samples shows good linearity.
Dedicated lifetime fits to ${B^+ \to K^+ \eta \gamma}$ samples yield 2.0$\pm$0.3~ps and 2.3$\pm$0.4~ps for $\etagg$ and $\etappp$, respectively. 
A lifetime fit to ${B^0 \to J/\psi K_S^0}$ using only $K_S^0$ to determine the signal vertex results in 1.528$\pm$0.027~ps.
A $CP$ asymmetry fit to the ${B^+ \to K^+ \eta \gamma}$ control samples yields (${\cal S}$,${\cal A}$)=(0.01$\pm$0.35, 0.06$\pm$0.29) and (0.2$\pm$0.6, 0.2$\pm$0.4) for $\etagg$ and $\etappp$, respectively.
Lastly, a $CP$ asymmetry fit to ${B^0 \to J/\psi K_S^0}$ only using $K_S^0$ to determine the signal vertex position yields (${\cal S}$,${\cal A}$)=(0.73$\pm$0.05, 0.00$\pm$0.03). 
These results are consistent with either their world-average or expected values~\cite{Patrignani:2016xqp}.

\section{Systematic Uncertainties}

We calculate systematic uncertainties in the following
categories by fitting the data with each fixed parameter
being varied by its uncertainty: values of physics parameters
such as $\Delta m_d$ and $\tau_{B^0}$, effective lifetime and $CP$ asymmetry
of the $B\bar{B}$ background, imperfect knowledge
of the $q\bar{q}$ background $\Delta t$ PDF, the flavor-tagging determination,
the signal and background fractions,
and the resolution functions. 
A possible bias in the fit is checked by performing a large number of pseudo-experiments.
The fit result is consistent with the input value within the statistical uncertainty. We quote this uncertainty as the possible fit bias. 
The uncertainty due to the vertex reconstruction is estimated by changing the requirements on the track quality.
For the effect of SVD misalignment, we use the value from the latest $\sin2\phi_1$ measurement at Belle~\cite{Adachi:2012et}, which is estimated from MC samples by artificially displacing the SVD sensors in a random manner.
Effects of tag-side interference~\cite{Long:2003wq} are estimated with a control sample of $B \to D^*\ell\nu$ events. 
A detailed description of the evaluation of the systematic uncertainties is found in Ref.~\cite{Santelj:2014sja}.
The dominant systematic contributions for ${\cal{S}}$ arise from the uncertainties in the resolution function and vertex reconstruction.
The systematic uncertainty in ${\cal{A}}$ is dominated by the resolution function.
These contributions
are added in quadrature and summarized in Table~\ref{tab:syst_total}.
\begin{table}[htb]
    \begin{center}
    \caption{Systematic uncertainties of ${\cal S}$ and ${\cal A}$.}
    \label{tab:syst_total}
    \begin{tabular}{ @{\hspace{0.5cm}}l@{\hspace{0.5cm}}|@{\hspace{0.5cm}}c@{\hspace{0.5cm}}c@{\hspace{0.5cm}} }
        \hline \hline
        Source                     & ${\cal S}$   & ${\cal A}$\\
        \hline
        Resolution parameters      &  $\pm$0.257  &  $\pm$0.049  \\
        Vertex reconstruction      &  $\pm$0.232  &  $\pm$0.022  \\
        Background $\Delta t$ PDF  &  $\pm$0.051  &  $\pm$0.006  \\
        Flavor tagging             &  $\pm$0.015  &  $\pm$0.019  \\
        Physics parameters         &  $\pm$0.004  &  $\pm$0.002  \\
        PDF for 3D fit             &  $\pm$0.096  &  $\pm$0.024  \\
        $CP$ violation in background &  $\pm$0.024  &  $\pm$0.022  \\
        Possible fit bias          &  $\pm$0.016  &  $\pm$0.015  \\
        Tag-side interference      &  $\pm$0.006  &  $\pm$0.010  \\
        \hline
        Total                      &  $\pm$0.364  &  $\pm$0.068  \\
        \hline \hline
    \end{tabular}
    \end{center}
\end{table}

\section{Confidence Level Contours}
\begin{figure}[htbp]
    \begin{center}
\ifprd
    \includegraphics[width=0.45\textwidth]{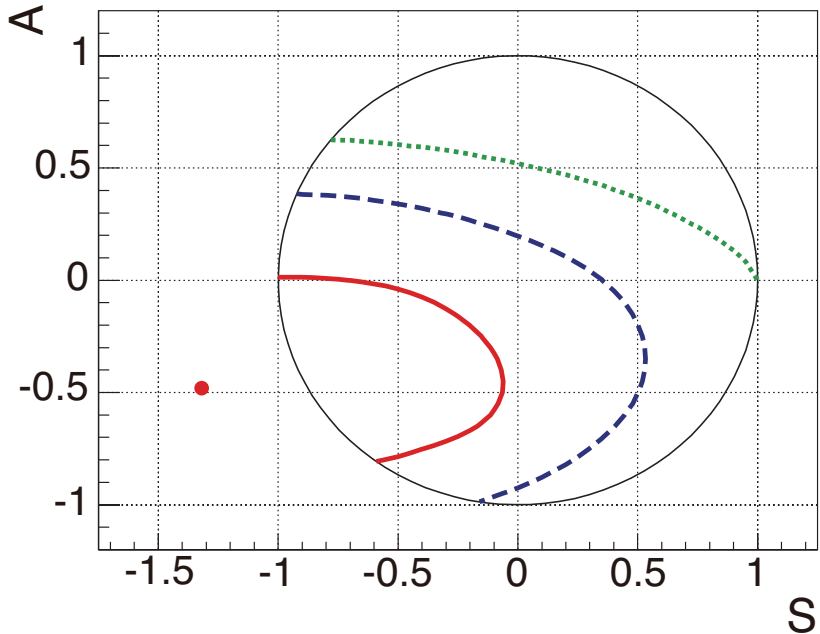}
\else
    \includegraphics[width=0.66\textwidth]{fig5.eps}
\fi
    \caption { The solid red, dashed blue and dotted green curves show the $1\sigma$, $2\sigma$ and $3\sigma$ confidence contours, respectively. The red dot shows the fit result. The physical boundary ${\cal{S}}^2 + {\cal{A}}^2 =1$ is drawn with a thin solid black curve. Our result is consistent with a null asymmetry within $2\sigma$.}
    \label{fig:contour}
    \end{center}
\end{figure}

Figure~\ref{fig:contour} shows confidence intervals calculated using the Feldman-Cousins frequentist approach~\cite{FeldmanCousins}, incorporating a smearing by additional Gaussian functions to represent the systematic uncertainties discussed above. Our result is less than $2\sigma$ away from zero, and is consistent with the BaBar result~\cite{Aubert:2008js} as well as the SM predictions~\cite{Grinstein:2004uu,Grinstein:2005nu,Matsumori:2005ax,Ball:2006cva,Ball:2006eu,Jager:2014rwa} with the assumption that time-dependent $CP$ asymmetries in ${B^0 \to K^{*0} \gamma}$ and ${B^0 \to K_S^0 \eta \gamma}$ are the same.

\section{Conclusion}

In summary, we have measured $CP$ violation parameters in $B^0 \to K_S^0 \eta \gamma$ decays
using a data sample of ${772\times 10^6 B\bar{B}}$ pairs.
The obtained parameters
\begin{eqnarray}
\nonumber
    {\cal S} &=& -1.32 \pm 0.77 {\rm (stat.)} \pm 0.36{\rm (syst.)}, \\
\nonumber
    {\cal A} &=& -0.48 \pm 0.41 {\rm (stat.)} \pm 0.07{\rm (syst.)}
\end{eqnarray}
are consistent with the null-asymmetry hypothesis within 2$\sigma$ as well as with SM predictions~\cite{Grinstein:2004uu,Grinstein:2005nu,Matsumori:2005ax,Ball:2006cva,Ball:2006eu,Jager:2014rwa}.
Our measurement is dominated by statistical uncertainty. 
Therefore, with much higher statistics and also higher acceptance and reconstruction efficiencies, the forthcoming Belle II experiment should significantly improve upon the precision of this measurement.

\section{Acknowledgments}

A.~I. is supported by the Japan Society for the Promotion of Science (JSPS) Grant No.~16H03968.
We thank the KEKB group for the excellent operation of the
accelerator; the KEK cryogenics group for the efficient
operation of the solenoid; and the KEK computer group,
the National Institute of Informatics, and the 
Pacific Northwest National Laboratory (PNNL) Environmental Molecular Sciences Laboratory (EMSL) computing group for valuable computing
and Science Information NETwork 5 (SINET5) network support.  We acknowledge support from
the Ministry of Education, Culture, Sports, Science, and
Technology (MEXT) of Japan, the Japan Society for the 
Promotion of Science (JSPS), and the Tau-Lepton Physics 
Research Center of Nagoya University; 
the Australian Research Council;
Austrian Science Fund under Grant No.~P 26794-N20;
the National Natural Science Foundation of China under Contracts
No.~11435013,  
No.~11475187,  
No.~11521505,  
No.~11575017,  
No.~11675166,  
No.~11705209;  
Key Research Program of Frontier Sciences, Chinese Academy of Sciences (CAS), Grant No.~QYZDJ-SSW-SLH011; 
the  CAS Center for Excellence in Particle Physics (CCEPP); 
Fudan University Grant No.~JIH5913023, No.~IDH5913011/003, 
No.~JIH5913024, No.~IDH5913011/002;                        
the Ministry of Education, Youth and Sports of the Czech
Republic under Contract No.~LTT17020;
the Carl Zeiss Foundation, the Deutsche Forschungsgemeinschaft, the
Excellence Cluster Universe, and the VolkswagenStiftung;
the Department of Science and Technology of India; 
the Istituto Nazionale di Fisica Nucleare of Italy; 
National Research Foundation (NRF) of Korea Grants No.~2014R1A2A2A01005286, No.2015R1A2A2A01003280,
No.~2015H1A2A1033649, No.~2016R1D1A1B01010135, No.~2016K1A3A7A09005 603, No.~2016R1D1A1B02012900; Radiation Science Research Institute, Foreign Large-size Research Facility Application Supporting project and the Global Science Experimental Data Hub Center of the Korea Institute of Science and Technology Information;
the Polish Ministry of Science and Higher Education and 
the National Science Center;
the Ministry of Education and Science of the Russian Federation and
the Russian Foundation for Basic Research;
the Slovenian Research Agency;
Ikerbasque, Basque Foundation for Science, Basque Government (No.~IT956-16) and
Ministry of Economy and Competitiveness (MINECO) (Juan de la Cierva), Spain;
the Swiss National Science Foundation; 
the Ministry of Education and the Ministry of Science and Technology of Taiwan;
and the United States Department of Energy and the National Science Foundation.

%


\begin{thebibliography}{99}
\bibitem{CC}
Throughout this paper, the inclusion of the charge conjugate mode decay is implied unless otherwise stated.
\bibitem{toyMC0}
We repeated the ensemble test with the input values of $( {\cal S}_{{\rm true}}, {\cal A}_{{\rm true}} )=(0.00, 0.00)$. The sensitivities 
are $( \delta {\cal S}, \delta {\cal A} )=(\pm0.59, \pm0.34)$ which are smaller than the uncertainties quoted in the text.
\end{thebibliography}
\end{document}